\documentclass[a4paper,11pt]{article}
\usepackage{jinstpub} 
\usepackage{siunitx}
\usepackage{subcaption}
\usepackage{float}
\usepackage{orcidlink}

\title{\boldmath Mechanism for reduction of the afterpulsing rate of PMTs}

\author[a]{Kai Morita\orcidlink{0009-0004-5641-5552},}
\author[b]{Mitsunari Takahashi\orcidlink{0000-0002-0574-6018},}
\author[a]{Habib Ahammad Mondal\orcidlink{0000-0001-7217-0234},}
\author[a]{Hidetoshi Kubo\orcidlink{0000-0001-9159-9853},}
\author[a]{Hideyuki Ohoka,}
\author[a]{Seiya Nozaki\orcidlink{0000-0002-6246-2767},}
\author[c]{Shunsuke Sakurai\orcidlink{0000-0001-7427-4520},}
\author[a]{Takayuki Saito\orcidlink{0000-0001-6201-3761},}
\author[d]{Tokonatsu Yamamoto,}
\author[a]{Yusuke Inome\orcidlink{0000-0003-2161-4469}}

\affiliation[a]{Institute for Cosmic Ray Research, The University of Tokyo,\\
5-1-5 Kashiwa-no-ha, Kashiwa, Chiba 277-8582, Japan}
\affiliation[b]{Institute for Space–Earth Environmental Research, Nagoya University,\\
Furo-cho, Chikusa-ku, Nagoya 464-8601, Japan}
\affiliation[c]{Osaka Metropolitan University,\\
3-3-138 sugimoto sumiyoshi-ku, Osaka 558-8585, Japan}
\affiliation[d]{Department of Physics, Konan University,\\
Kobe, Hyogo 658-8501, Japan}

\emailAdd{morikai@icrr.u-tokyo.ac.jp}

\abstract{
Photomultiplier tubes (PMTs) are used in Imaging Atmospheric Cherenkov Telescopes (IACTs) to detect Cherenkov light produced by air showers induced by gamma rays in the atmosphere. 
The afterpulsing rate of the PMTs for the Large-Sized Telescopes (LSTs) of the Cherenkov Telescope Array Observatory (CTAO) was found to increase if they were kept unused in storage.
In contrast, PMTs that had been operated in the first LST showed a slight decrease in the rate. 
This decrease could be explained by a reduction of residual gas caused by ion feedback, although the detailed mechanism remained unclear.
In this study, to investigate factors responsible for the evolution in the afterpulsing rate, we operated several PMTs under different high voltage and light illumination conditions. 
We monitored their rate daily for three weeks to compare their evolution under different conditions. 
We found that the reduction of afterpulses require both illumination and high-voltage operation.
Notably, the reduction strongly depends on the applied high voltage and is closely correlated with the integrated anode current.
Therefore, we conclude that the reduction of residual gas is mainly caused by ionization occurring at later dynodes of the PMTs, and the ions are trapped by the dynodes.
We also discuss a possible explanation of the reduction of afterpulsing rate by later dynodes.
}

\keywords{Electron multipliers (vacuum), Ionization and excitation processes, Vacuum-based detectors, Cherenkov detectors, Gamma telescopes}

\begin{document}
\maketitle
\flushbottom

\section{Introduction}
\label{sec:intro}
    Photomultiplier tubes (PMTs) are highly sensitive photon detectors and are widely used in very-high-energy ($\gtrsim \SI{100}{GeV}$) gamma-ray astronomy.
    Imaging Atmospheric Cherenkov Telescopes (IACTs), which are based on the ground, detect Cherenkov light produced by air showers induced by cosmic gamma rays. 
    Since PMTs are exposed to the night sky during observations, continuous illumination by background light~\citep{NSB_LaPalma} results in high photoelectron rates and frequent afterpulsing (AP).
    Moreover, the AP rate increases over time due to helium permeation from the atmosphere through the photocathode~\citep{coates1973origins:IonTrap,incandela1988performance}. This may deteriorate the telescope performance.

    Cherenkov Telescope Array Observatory (CTAO)
    is a next-generation IACT. Its Large-Sized Telescopes (LSTs)~\citep{CTAO:2025:PMTmodules} and Medium-Sized Telescopes~\citep{Tsiahina:2021:NectarCAM,WERNER201731} also employ PMTs, which were developed together by HAMAMATSU PHOTONICS K.K..
    The authors of~\citep{Takahashi:2025:PD24} measured a change in the AP rate of those PMTs. They found that the rate in spare PMTs kept in storage increases over time, but such a trend was not observed in PMTs used in the first LST.
    In addition, laboratory measurements confirmed that the AP rate decreases when the PMTs are operated under light illumination and high voltage (HV) applied, as is the case for PMTs in the telescope during observations.

    Such a decrease could be explained by the removal of residual gas from the vacuum caused by ionization.
    Accelerated photoelectrons ionize residual molecules --- the same process that underlies the AP generation~\citep{Morton1967}. The resulting ions drift towards the photocathode, where some are trapped into the metal.
    However, it remained unclear whether the observed reduction requires both light illumination and HV operation, and whether the photoelectron current dominates the ionization process.
    To address these questions, we performed a controlled experiment in the laboratory.

\section{Method}
\label{sec:method}
\subsection{Long-term monitoring}
\label{subsec:long-term}

\pagebreak
    To confirm the reduction of the AP rate associated with ionization processes inside the PMT, we kept 21 PMTs in different conditions for approximately three weeks. 
    These PMTs are R12992-100~\citep{CTA:2017:PMTevaluation}, which were manufactured for the second to fourth LSTs, and we used their spares as test samples.
    
    They were put in a dark box and illuminated by an LED while the photocathodes of some of them were masked to block the light. In addition, different values of the HV including \SI{0}{\volt} were applied. 
    The potential difference between the photocathode and the first dynode is fixed at \SI{350}{\volt} regardless of the applied voltage provided it is non-zero.
    The LED provided diffused and temporally constant light corresponding to a photoelectron rate of $\sim$ \SI{2.5}{\giga\hertz} --- an order of magnitude higher than the typical LST night-sky background.
    These conditions are summarized in Table~\ref{tab:i}.
\begin{table}[htbp]
\centering
\caption{PMT conditions for the long-term operation.\label{tab:i}}
\smallskip
\begin{tabular}{c|c|c|c}
\hline
PMT group & Light & High voltage [V] & Number of PMT\\
\hline
Group 1 & Illuminated & 1100 & 5\\
Group 2 & Illuminated & 750 & 5\\
Group 3 & Illuminated & 0 & 5\\
Group 4 & Masked & 1100 & 3\\
Group 5 & Masked & 0 & 3\\
\hline
\end{tabular}
\end{table}
    
    Throughout the three weeks, we monitored the AP rate at least once per day.
 Here, the AP rate is defined as the number of afterpulses divided by the number of incident photoelectrons.
    To obtain the two numbers, we shot all the PMTs with a HV of \SI{1100}{\volt} by laser pulses with a sub-ns width~\citep{Inome:2017:Laser}. The waveform was taken with a trigger synchronized to the laser, and this enabled us to find a main pulse produced by photoelectrons and subsequent afterpulses separately.

    If the same ionization process is responsible for both the afterpulses and their reduction, the same HV dependence should appear in them. 
    To clarify it, we also measured and compared the AP rates at \SI{1400}{\volt} and at \SI{1100}{\volt} for the same set of 21 PMTs before the long-term operation began.

    For data acquisition and control, seven PMTs together with a single readout board form one PMT module~\citep{CTAO:2025:PMTmodules}.
    Its readout system is capable of acquiring \SI{1}{\micro\second} waveforms with \SI{1}{\giga\hertz} sampling.
    We used these modules to continuously monitor the anode current of each PMT.
   
\subsection{Gain measurements}
\label{subsec:gain measurement}
    To convert the charge of the main pulse to the photoelectron number, we measured the PMT gain.
    Since the gain may vary, we did it prior to each daily afterpulse measurement. 
    The measured value is also used for two other purposes. First, we evaluated the charge of each afterpulse in a photoelectron-equivalent number to compare other results measured with a different gain. 
    Second, we monitored the cathode current and the photoelectron incident rate by dividing the anode current by the gain. 
    If the residual gas is ionized between the photocathode and the first dynode (hereafter the front region), the AP rate reduction should be proportional to them.
    
    The gain was measured by the output charge distribution in response to single photoelectron inputs.
    First, we measured the gain at \SI{1400}{\volt}, a sufficiently high HV to resolve single photoelectrons. 
    Ten thousand waveforms with a duration of \SI{100}{\nano\second} were acquired at \SI{1400}{\volt}. 
    The single-photoelectron pulse has a mean full width at half maximum (FWHM) of $\sim2.3$ ns at 1400 V.
    Therefore, the ADC counts in each waveform were summed within a 3 ns sliding window,
    which is sufficient to contain the SPE pulse while optimizing the signal-to-noise ratio, and the highest sum was derived.
    The extracted values are filled into histogram, and the noise component and the single photoelectron one were fitted with a double Gaussian function. 
    The gain was then derived from the difference between the peak values.

    Since the signal-to-noise ratio of single photoelectron signals is insufficient at 1100 V and 750~V, which were the HV settings used during the long-term operation, the gain at these voltages was derived by multiplying the value at \SI{1400}{\volt} by the gain ratio between these voltages.
    To determine the ratio, relatively intense laser pulses were shot into the PMTs.
    The mean FWHM is $\sim$2.7 ns at 1100 V and $\sim$3.4 ns at 750 V.
    Therefore, for each waveform, the ADC counts around the peak were integrated over a \SI{5}{\nano\second} window to fully contain the pulse.
    The resulting integrated ADC counts were then averaged over all events for each voltage, and the ratio of this average to that obtained at \SI{1400}{\volt} was taken.

\subsection{Afterpulse measurements}
\label{subsec:ap measurement}
    Laser pulses corresponding to $\sim$50 photoelectrons were shot on the PMTs, and 100,000 waveforms with a duration of \SI{1}{\micro\second} were acquired.
    The HV was set to \SI{1100}{\volt} regardless of the value applied during the long-term operation.
    On one hand, we estimated the number of photoelectrons in the main pulse produced by the laser pulse by summing ADC counts over \SI{5}{\nano\second} around the peak, in the same manner as in Section~\ref{subsec:gain measurement}.
    By dividing this value by the gain at \SI{1100}{\volt} and the other constants,
    the number of photoelectrons in the main pulse was calculated.
    On the other hand, the number of afterpulses was evaluated as follows.   
    The ADC counts were summed over every \SI{3}{\nano\second} sliding window after the main pulse in the 100,000 waveforms.
    Only if this sum exceeded the 4 photoelectron equivalent charge, the signal was counted as an afterpulse to reject electrical or stray-light noise.  

\section{Results and Discussion}
\label{sec:results and discussion}
\subsection{Reduction of residual gas}
\label{subsec:reduction}
\begin{figure}[htbp]
\centering

\begin{subfigure}[t]{0.49\textwidth}
    \centering
    \includegraphics[width=\textwidth]{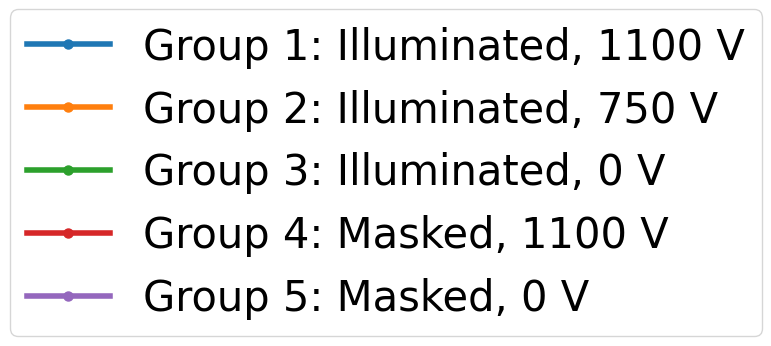}
\end{subfigure}
\hfill
\begin{subfigure}[t]{0.49\textwidth}
    \centering
    \includegraphics[width=\textwidth]{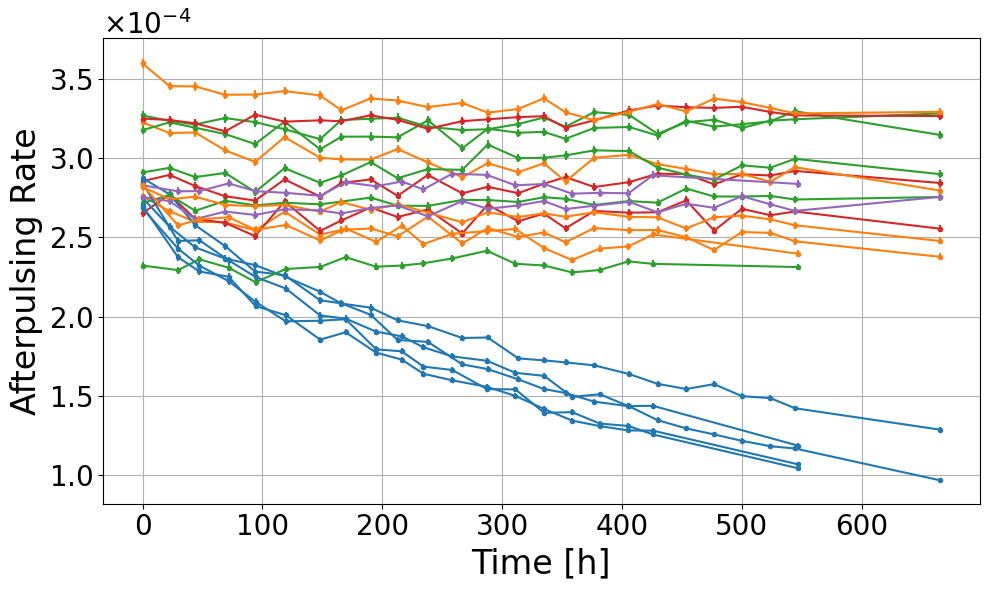}
    \caption{}
\end{subfigure}

\vspace{0.8em}


\begin{subfigure}{0.49\textwidth}
    \centering
    \includegraphics[width=\textwidth]{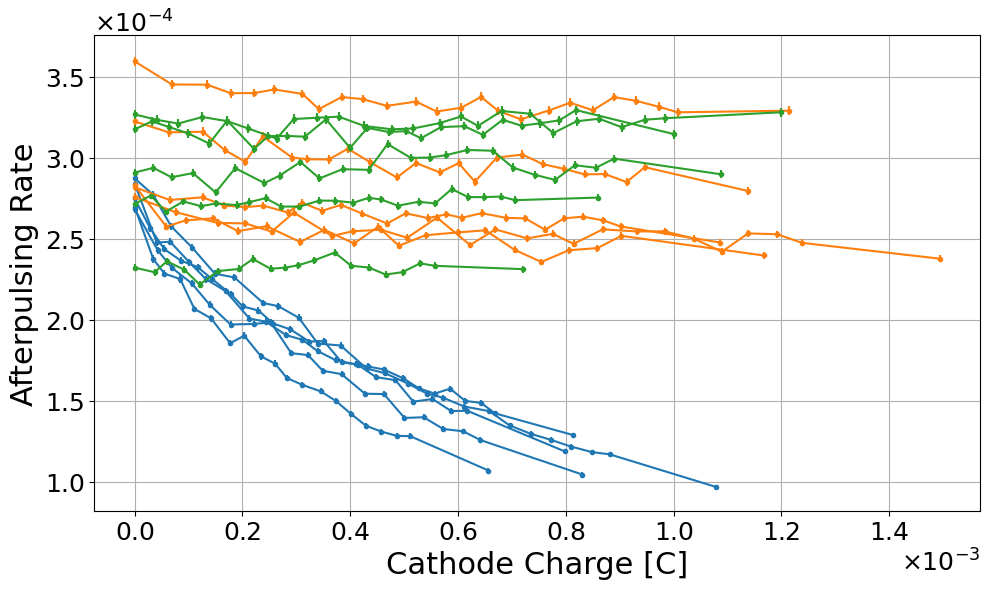}
    \caption{}
\end{subfigure}
\hfill
\begin{subfigure}{0.49\textwidth}
    \centering
    \includegraphics[width=\textwidth]{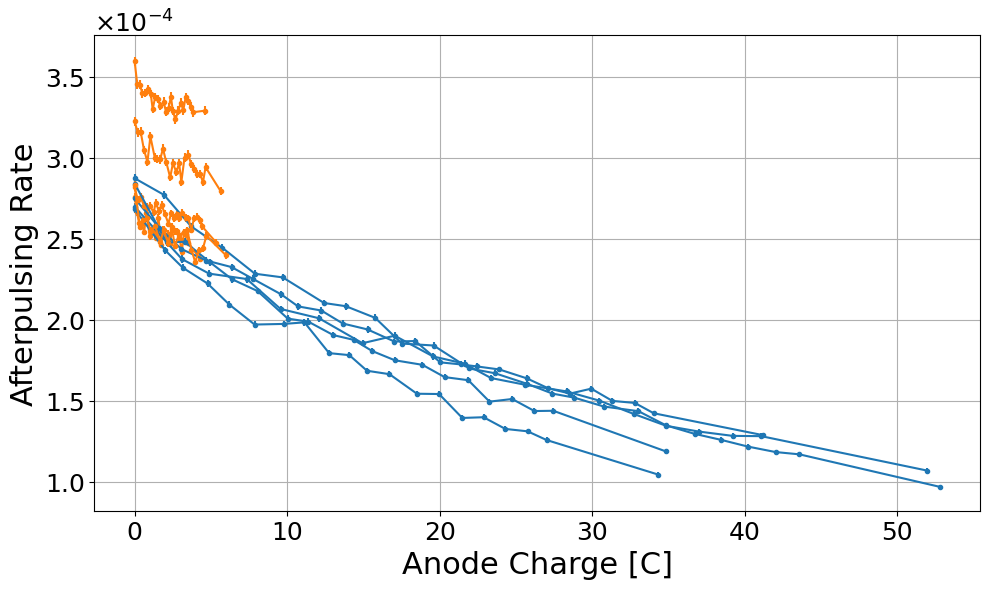}
    \caption{}
\end{subfigure}
\caption{Evolution of the AP rate during the long-term operation
(a) Time evolution of the rate for all 21 PMTs. 
(b) AP rate as a function of the integrated cathode current of the 15 PMTs illuminated during the long-term operation. 
(c) AP rate as a function of the integrated anode current of the 10 PMTs operated with both illumination and high voltage.
}
\label{fig:ap_reduction}
\end{figure}

    The PMTs illuminated and a HV of \SI{1100}{\volt}  applied exhibited a clear decline in the AP rate as shown by the blue line in panel (a) of figure~\ref{fig:ap_reduction}. 
    In contrast, PMTs without illumination and/or without HV (green, red, and purple lines) did not show such a decline.
    These results indicate that the reduction of the AP rate requires both illumination and high-voltage application. 
    This supports the idea that the afterpulses are reduced by accelerated electrons ionizing the residual gas inside the PMT.

    The change in the AP rate with the integrated cathode current (converted to charge) is shown in panel (b) to inspect the ionization in the front region.
    We calculated the cathode charge for the PMTs operated under illumination at HV, using the anode charge and the measured gain.
    For the PMTs illuminated without HV application (green line), the cathode charge was estimated using the ratio of the incident photoelectron rate of these PMTs to the illuminated ones. The ratio was measured with HV application before the long-term operation. 
    The decrease in the AP rate for the PMTs operated at \SI{1100}{\volt} (blue line) is much larger than that at \SI{750}{\volt} (orange line) for the same cathode charge.
    As described above, the potential difference between the photocathode and the first dynode is fixed at \SI{350}{\volt} regardless of the HV setting.
    Therefore, if the ionization responsible for the AP rate reduction were mainly caused by photoelectrons produced at the photocathode, such a difference in the decrease would not be expected.
    This result clarifies that the reduction is not mainly due to the photoelectrons produced at the photocathode.

    The change in the AP rate with the anode charge is shown in panel (c). 
    The reduction in the AP rate for PMTs operated at \SI{750}{\volt} (orange line) is comparable to that for PMTs at \SI{1100}{\volt} (blue line) having the same anode charge.
    This indicates that the ionization responsible for this reduction mainly occurs in the region among the later dynodes and the anode (hereafter the rear region).
    To quantify the rate of reduction, we performed a linear fit for each condition using the function $R_{\mathrm{AP}} = a\,Q + b$,
    where $R_{\mathrm{AP}}$ is the AP rate and $Q$ is the anode charge in coulombs [C].
    For the fitting, all data points were used for 750 V, while only the first four data points were used for 1100 V.
    We obtained a slope
    $a = (-7.15 \pm 0.20)\times10^{-6}\ \mathrm{C^{-1}}$
    for the PMTs operated at \SI{1100}{\volt}, and
    $a = (-4.50 \pm 0.15)\times10^{-6}\ \mathrm{C^{-1}}$
    for those operated at \SI{750}{\volt}.
    The magnitude of the slope at \SI{1100}{\volt} is approximately 1.6 times larger than that at \SI{750}{\volt}.
    The difference in the reduction amount is attributed to the difference in the ionization cross section between \SI{1100}{\volt} and \SI{750}{\volt}.
    At \SI{1100}{\volt}, the potential difference between dynodes is approximately \SI{100}{\volt}, whereas it is about \SI{60}{\volt} at \SI{750}{\volt}. 
    When electrons are accelerated up to energies of roughly \SI{100}{\electronvolt} in the former case and \SI{60}{\electronvolt} in the latter, the ionization cross section for helium, which dominates AP of R12992-100, differs by a factor comparable to the difference in the measured slopes~\citep{Shah:1988:Ionization}.
    Consequently, the decrease in the AP rate is likely explained by a combination of the number of secondary electrons and their energy.

\subsection{Region responsible for the afterpulses}
\label{subsec:ion}
\begin{figure}[htbp]
\centering
\includegraphics[width=.49\textwidth]{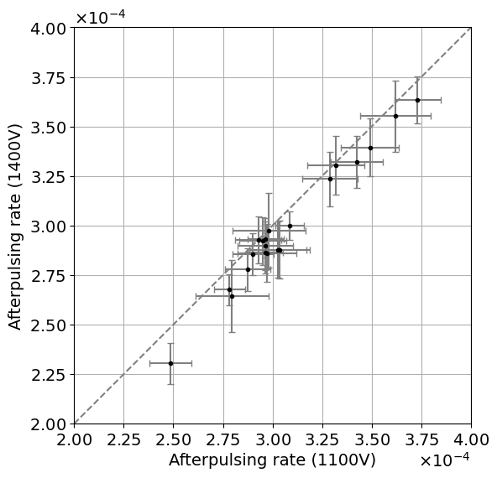}
\caption{
AP rate of the 21 PMTs measured at \SI{1400}{\volt} against that at \SI{1100}{\volt}. The cross bars represent a sum of the statistical standard error and a systematic error coming from the measured PMT gain.
}
\label{fig:APvsHV}
\end{figure}

    As discussed in section~\ref{subsec:reduction}, a larger gain implies more frequent ionization. 
    To investigate whether this dependence also appears in the AP rate, we compared it under two voltages---\SI{1400}{V} and \SI{1100}{V}.
    As shown in figure~\ref{fig:APvsHV}, the rate is almost the same regardless of the voltage applied during the measurements, whereas more ions are expected to be produced at \SI{1400}{V} than at \SI{1100}{V}.
    This paradox is explained as follows. Most of the ions produced at the rear region do not reach the photocathode and cannot contribute to AP. They are instead neutralized and captured by the dynode surfaces, as discussed in ref.~\citep{coates1973origins:IonTrap}. The observed afterpulses are considered to originate mainly in the front region, where the electric field is fixed regardless of the total HV.

\section{Conclusion}
\label{sec:conclusion}
    We confirmed that light illumination combined with HV operation is necessary for the AP reduction. 
    This induces ionization inside the PMT, leading to the removal of residual gas. 
    We also found that the decrease in the rate strongly depends on the applied HV through the PMT gain.
    This result indicates that ionization is mainly caused by secondary electrons generated and multiplied at the later dynodes. 
    The resulting depletion at the later stages drives diffusion from the front region to the rear, and the density is homogenized.
    Since afterpulses mainly originate from the front region, this diffusion decreases the AP rate.

\acknowledgments
    This work was supported by the Japan Society for the Promotion of Science through KAKENHI Grants 23H04897, 21H04468, and 23H05430.
    We also thank HAMAMATSU PHOTONICS K.K. for the fruitful discussions with us.




\bibliographystyle{aux/JHEP}
\bibliography{biblio.bib}




\end{document}